\documentclass[10pt,aps,
notitlepage,%
oneside,%
twocolumn,
nofootinbib,%
superscriptaddress,%
floatfix, pra]%
{revtex4-2}
\pdfoutput=1
\usepackage{graphicx} 
\usepackage[top=3cm, bottom=3cm, left = 2cm, right = 2cm]{geometry} 
\usepackage[utf8]{inputenc}
\usepackage{textcomp,soul}
\usepackage{siunitx}
\usepackage{subcaption}
\usepackage[dvipsnames]{xcolor}
\usepackage{braket}
\usepackage{amsmath,amssymb}  
\usepackage{esvect}
\usepackage{bm}  
\usepackage{mathtools}
\usepackage{caption}
\usepackage[pdftex,bookmarks,colorlinks,breaklinks]{hyperref} 
\usepackage{float}
\usepackage{graphicx,psfrag,rotating,epstopdf}

\usepackage{bm}
\usepackage{tikz}

\usepackage{memhfixc} 
\usepackage{fancyhdr}
\usepackage{amssymb}
\usepackage{ragged2e} 

\begin{document}
\title{Magnetometry with Broadband Microwave Fields in Nitrogen-Vacancy Centers in Diamond}

\author{Arezoo Afshar}
\affiliation{National Research Council of Canada, 100 Sussex Drive, Ottawa, Ontario K1N 5A2, Canada}
\affiliation{\mbox{Department of Physics, University of Ottawa, 25 Templeton Street, Ottawa, Ontario, K1N 6N5 Canada}}
\author{Andrew Proppe}
\affiliation{National Research Council of Canada, 100 Sussex Drive, Ottawa, Ontario K1N 5A2, Canada}
\affiliation{\mbox{Department of Physics, University of Ottawa, 25 Templeton Street, Ottawa, Ontario, K1N 6N5 Canada}}
\author{Noah Lupu-Gladstein}
\affiliation{National Research Council of Canada, 100 Sussex Drive, Ottawa, Ontario K1N 5A2, Canada}
\affiliation{\mbox{Department of Physics, University of Ottawa, 25 Templeton Street, Ottawa, Ontario, K1N 6N5 Canada}}
\author{Lilian Childress}
\affiliation{Physics Department, McGill University, 3600 Rue University, Montr\'eal QC H3A 2T8, Canada}
\author{Aaron Z. Goldberg}
\affiliation{National Research Council of Canada, 100 Sussex Drive, Ottawa, Ontario K1N 5A2, Canada}

\author{Khabat Heshami}
\affiliation{National Research Council of Canada, 100 Sussex Drive, Ottawa, Ontario K1N 5A2, Canada}
\affiliation{\mbox{Department of Physics, University of Ottawa, 25 Templeton Street, Ottawa, Ontario, K1N 6N5 Canada}}

\date{\today}
%
\begin{abstract}
    Nitrogen-vacancy (NV) centers in diamond are optically addressable and versatile light-matter interfaces with practical application in magnetic field sensing, offering the ability to operate at room temperature and reach sensitivities below pT/$\sqrt{\mathrm{Hz}}$. We propose an approach to simultaneously probe all of the magnetically sensitive states using a broadband microwave field and demonstrate that it can be used to measure the external DC magnetic field strength with sensitivities on the order of 10~pT/$\sqrt{\mathrm{Hz}}$. We develop tools for analyzing the temporal signatures in the transmitted broadband microwaves to estimate the magnetic field, comparing maximum likelihood estimation based on minimizing the Kullback-Leibler divergence to various neural network models, and both methods independently reach practical sensitivities. 
    These results are achieved without optimizing parameters such as the bandwidth, 
and shape of the probing microwave field such that further improvements in sensitivity can be envisioned. Our results motivate novel implementations of NV-based magnetic sensors with the potential for vectorial magnetic field detection at 1-10 kHz update rates and improved sensitivities without requiring a bias magnetic field.
\end{abstract}
\maketitle
\section{Introduction}
Color centers in diamond provide an attractive light-matter interface for applications in quantum sensing~\cite{degen_quantum_2017} and development of quantum networks~\cite{doherty2013nitrogen,bernien2013heralded,hensen2015loophole,nguyen2019quantum}. Nitrogen-vacancy (NV) centers in diamond have become a well-established system for magnetic sensing~\cite{degen2008scanning, taylor2008high, maze2008nanoscale, rondin2014magnetometry, barry2020sensitivity} owing to their long spin coherence times~\cite{bar2013solid}, optical spin preparation, and a variety of spin detection techniques~\cite{hopper2018spin}. The NV center's general versatility and ability to operate from cryogenic to above room temperature~\cite{toyli2012measurement} have been key to applications ranging from nanoscale materials imaging~\cite{hong2013nanoscale, rovny2024nanoscale} to neuroscience~\cite{barry2016optical}.

The four vectorial orientations of NV defects and the hyperfine coupling to nearby nuclear spins leave a rich spectrum of resonances in the microwave (MW) domain, whose frequencies encode information about the vector components of the DC magnetic field. Scanning these spin resonances with a tunable MW field and off-resonant optical readout is the basis of the widely employed optically detected magnetic resonance (ODMR) modality for magnetic field sensing~\cite{gruber1997scanning, maertz2010vector}. To more efficiently probe multiple NV orientations, vector magnetometers employ sequential measurements of different resonances~\cite{clevenson2018robust} or complex multi-frequency lock-in techniques for simultaneous detection of all orientations~\cite{schloss2018simultaneous}. While fluorescence detection of the NV spin is most common, other methods such as photoelectric readout~\cite{bourgeois2015photoelectric, siyushev2019photoelectrical}, spin-to-charge conversion~\cite{shields2015efficient, hopper2016near-infrared-assisted}, IR absorption~\cite{acosta2010broadband}, or laser-threshold~\cite{jeske2016laser} are also possible, in each case detecting the spin populations without directly probing coherent processes on the spin transitions. In contrast, magnetic sensing based on electromagnetically induced transparency~\cite{acosta2013electromagnetically} or MW cavity coupling~\cite{ebel2021dispersive, eisenach2021cavity-enhanced, wang2024spin-refrigerated} exploits the coherent interactions of the NV spins with incident electromagnetic radiation. Aided by low shot noise power in the MW regime, MW cavity-coupling approaches~\cite{wang2024spin-refrigerated} have reached sensitivities rivaling the best fluorescence detection results~\cite{barry2024sensitive}. 
However, narrowband high-finesse MW cavities probe only a limited spectral window at a time, which restricts simultaneous access to all NV transitions and can limit dynamic range under slowly varying background fields. In contrast, our broadband-pulse approach spans the entire NV ground-state resonance manifold, addressing all crystallographic orientations and hyperfine transitions simultaneously. As the background magnetic field varies, the resonance frequencies shift but remain within the spectral window of the probe pulse, resulting in a smooth and uniquely identifiable change in the transmitted time-domain waveform. 
Here, we consider broadband MW probes for NV magnetometry, theoretically examining the transmission of a short-duration MW pulse through an NV ensemble. In particular, the transmitted pulse encodes the full spectrum of NV spin resonances into its temporal response, which can be directly measured by fast MW receivers without the need for frequency-domain spectroscopy. The primary challenge that we broach in this paper is how to accurately and robustly invert such a time-domain signal to extract the magnetic field. While we consider the field magnitude in this initial study, this approach permits sampling all orientations simultaneously, offering a clear path towards high-bandwidth vector magnetometry with shot-noise-unlimited MW detection. Without fully optimizing our protocol, we demonstrate that analyzing the transmitted MW signal in the time domain is sufficient to reach sensitivities that are comparable to those of ODMR. In addition, our approach remains accurate at small magnetic field values without requiring a bias field to spectrally separate resonances, a feature that may further improve practical applications of these sensors.

The remainder of the paper is structured as follows: We begin by describing our approach, where we use the NV center's electronic and nuclear spin Hamiltonian to find an expected absorption profile in the MW domain, then use the Kramers-Kronig relation to determine the full response function for evaluation of the transmitted MW signal for any given external magnetic field. In Section~\ref{results}, we first use minimization of Kullback-Leibler (KL) divergence curves to find our best estimate of the magnetic field for a given output signal. We also discuss an independent approach based on neural networks to extract external magnetic field strengths from the output MW signal. Finally, in Section~\ref{sensitivity}, we evaluate and discuss the sensitivity of our approach based on these techniques.

\section{Scheme}
\begin{figure}[htbp]
    \centering
    \includegraphics[width=1\columnwidth, viewport={7 1 850 460}, clip] {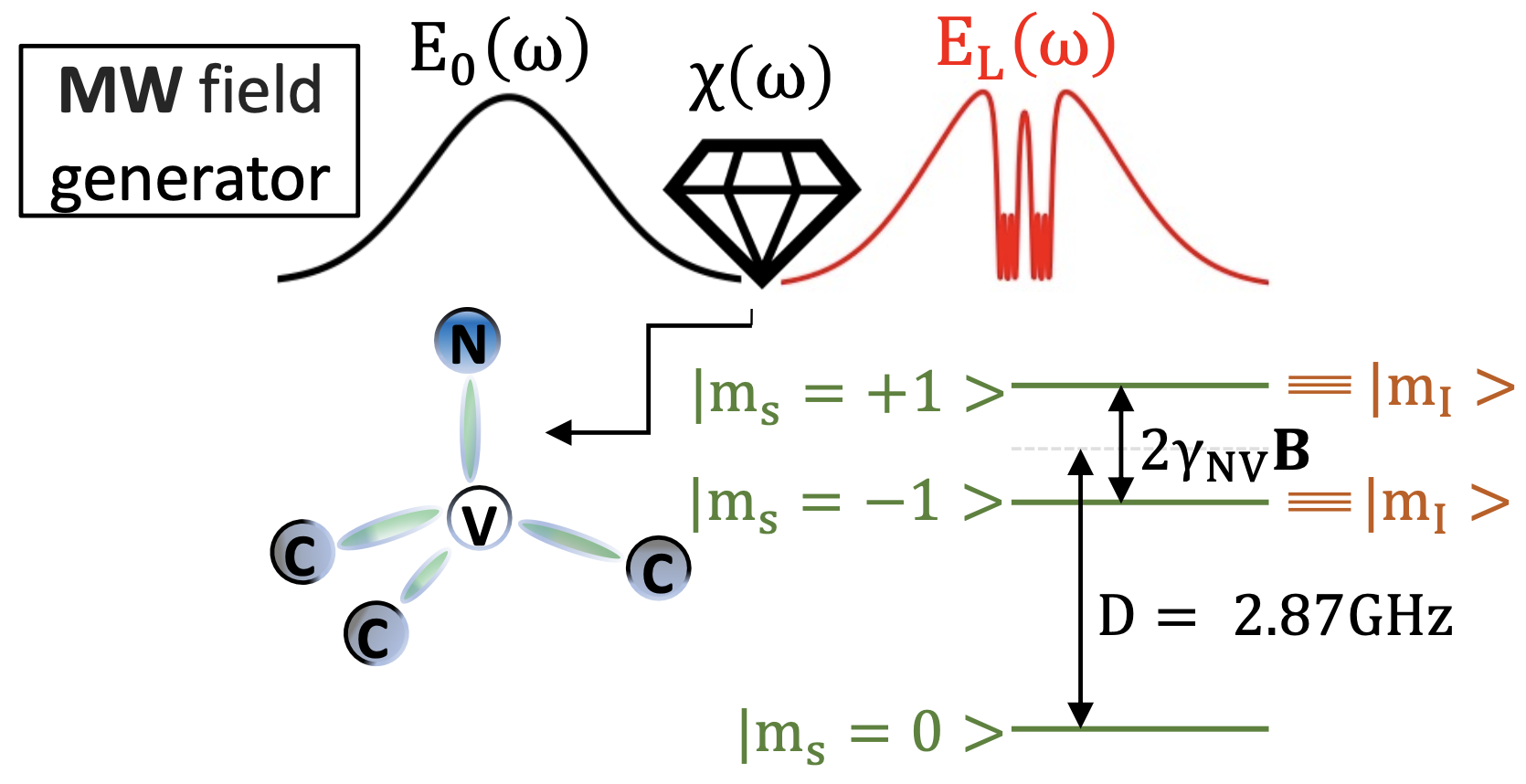}
    \caption{\justifying In this scheme, the MW field generator sends a broadband pulse to the NV ensemble. This can be achieved with a stripline waveguide, allowing evanescent coupling of the MW field to the NV centers or NV ensemble embedded in a MW cavity. As the MW input propagates through the sample, the NV centers partially absorb the signal and consequently modify the signal's intensity. This alteration is depicted in the output signal (red curve). The external magnetic field influences the frequencies at which NV centers absorb the microwave field. This leads to changes in the partially transmitted MW signal, thereby providing information about the underlying magnetic field. Shown below are the atomic and quantum structures of the NV center.}
    \label{fig:scheme}
\end{figure}

In this work, we aim to develop a technique for magnetic field sensing that simultaneously addresses multiple transitions between different spin states of NV centers in the ground state manifold using a broadband MW field. This approach could potentially reveal more information from the spectrum and, consequently, provide a more accurate measurement of the magnetic field. As depicted in Fig.~\ref{fig:scheme}, all magnetically sensitive states of an NV center in diamond are probed with a broadband MW field. At high NV concentrations, this will result in an absorption signature on the transmitted MW field that will be dependent on the external magnetic field. 

Inspired by methods in inhomogeneous absorption engineering for photon echo quantum memories for light~\cite{afzelius2009multimode,tittel2010photon,heshami2016quantum}, we expect the output signal to carry information about the position of the absorption peaks and therefore the external magnetic field. For example, a periodic absorption peak separation of $\delta$ (as proposed in atomic frequency comb quantum memories~\cite{afzelius2009multimode}) will result in a echos in the time domain that appear at the intervals of $2\pi/\delta$. For this approach to be applicable to any given magnetic field, we need to develop methods to analyze the output MW signal in the time domain. This will be covered later in this section.

%
\par We begin by describing the spin structure of a negatively charged NV center, which comprises an $\mathbf{S} = 1$ electronic spin and an $\mathbf{I} = 1$ nuclear spin associated with the host $^{14}$N nitrogen. The state space is spanned by the states $\ket{m_{s}, m_{I}}$, where $m_s$ and $m_I$ are the electronic and nuclear spin Zeeman sublevels, respectively.

\begin{figure}
    \centering
    \includegraphics[width=1\columnwidth, viewport={8 5 570 420}, clip]{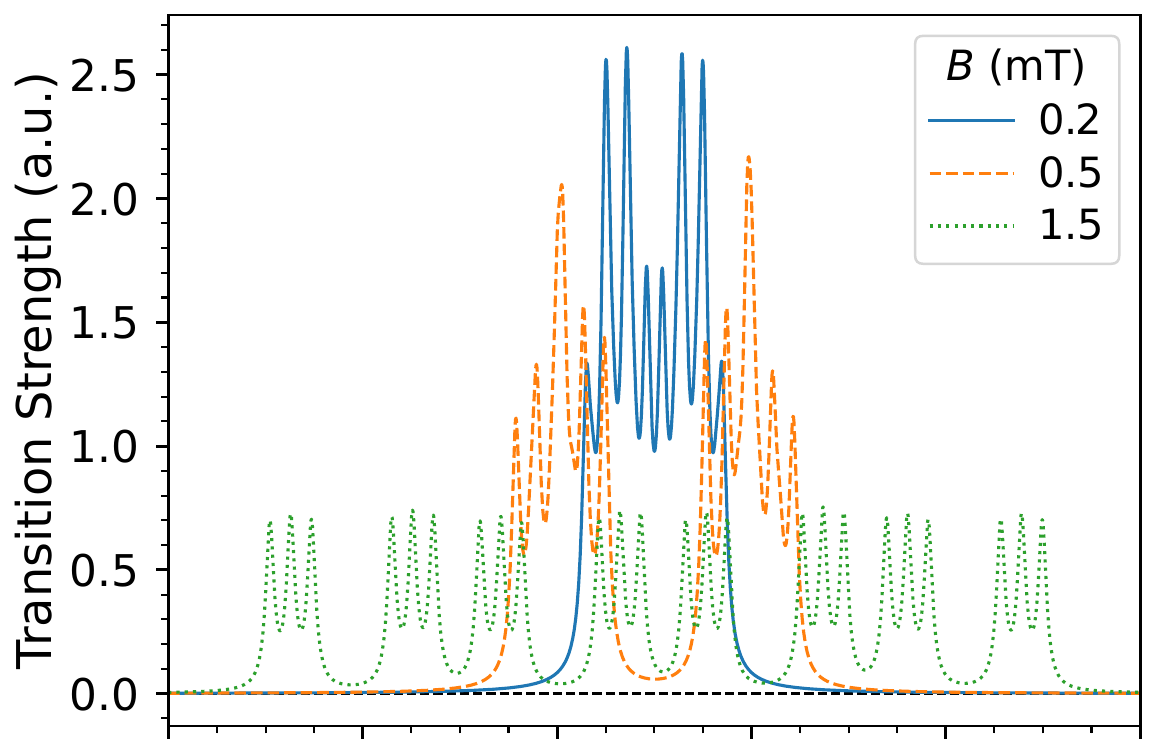}
     \includegraphics[width=1\columnwidth, viewport={8 3 570 420}, clip]{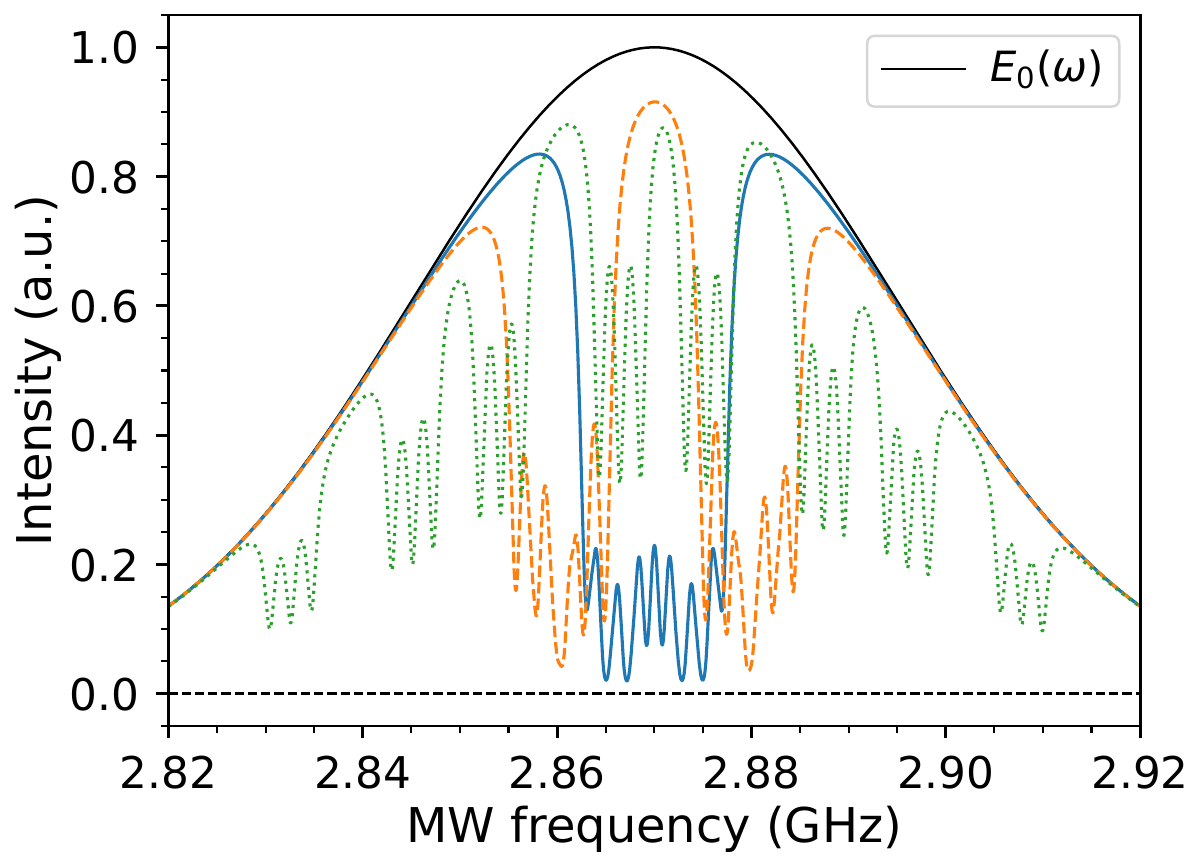}
    \caption{\justifying Top: Expected MW absorption in an ODMR measurement, due to spin transitions from $|m_{s} = 0\rangle$ to $|m_{s} = \pm 1\rangle$ in ensemble NV centers, are given by the ground-state-manifold Hamiltonian and combined with Lorentzian lines with 1 MHz linewidth. Magnetic field is oriented along angular coordinates $(\pi/6,\pi/3)$ relative to the diamond.
    Bottom: Fourier transform of a Gaussian pulse (85 MHz FWHM) transmitted through an ensemble of NV centers in different magnetic fields. In the frequency domain and within linear response theory, the absorption profile from the top graph is imprinted on the MW spectrum.
    }
    \label{fig:ODMR_broadbandaddresing}
\end{figure}

The total Hamiltonian for the NV center can thence be broken into three contributions:
\begin{equation}\label{eq:ground_state_hamiltonian}
\hat{H}_{total} = \hat{H}_{el} + \hat{H}_{nuc} + \hat{H}_{el-nuc}.
\end{equation}
When an external magnetic field $\textbf{B}$ is applied, due to the Zeeman effect, the degeneracy of the $\ket{m_{s} = \pm 1}$ levels lifts by up to $2\gamma_{e}B$ and $\hat{H}_{el}$ becomes:
\begin{equation}\label{eq:electronic_hamiltonian}
\frac{1}{h}\hat{H}_{el} = D \hat{S}^{2}_{z} + \gamma_{e}\textbf{B}\cdot\hat{\textbf{S}},
\end{equation}
where $D = 2.87$~GHz is the temperature-dependent zero field splitting along the NV symmetry axis $\hat{z}$~\cite{cambria_physically_2023}, and $\gamma_{e}$ is the gyromagnetic ratio.
The $^{14}N$ nuclear spin Hamiltonian in the presence of an external magnetic field is:
\begin{equation}\label{eq:nuclear_hamiltonian}
\frac{1}{h}\hat{H}_{nuc} = P \hat{I}^{2}_{z} + \gamma_{n}\textbf{B}\cdot\hat{\textbf{I}}.
\end{equation}
The quadrupole component $P = -4.95$~MHz ~\cite{Smeltzer2009Robust} lowers the energies of the states $\ket{m_{I} = \pm 1}$ below that of $\ket{m_{I} = 0}$, and $\gamma_{n}$ is the nuclear gyromagnetic ratio. Furthermore, hyperfine interactions generate an additional nuclear-spin-dependent splitting in the ground state manifold through
\begin{equation}\label{eq:spin_nuclear_hamiltonian}
\frac{1}{h}\hat{H}_{el-nuc} = A_{\parallel} S_{z}I_{z} + A_{\perp}(S_{x}I_{x} + S_{y}I_{y}).
\end{equation}
Here, $A_{\parallel} = 2.16$~MHz~\cite{Smeltzer2009Robust} and $A_{\perp} = 2.7$~MHz~\cite{felton_hyperfine_2009} are the magnetic hyperfine parameters parallel and perpendicular to the NV center axis, respectively. 
Transitions are driven when the NV center is subjected to a MW field resonant with the energy difference between \(m_s = 0\) and \(m_s = \pm 1\). (Direct \(-1 \leftrightarrow +1\) transitions are dipole-forbidden and negligibly weak in the magnetic dipole approximation.) The electron spin transitions under different magnetic fields are shown in Fig.~\ref{fig:ODMR_broadbandaddresing}(a). These are assumed to have Lorentzian absorption lines \cite{zhu2023simulation}. 


As described above and shown in Figs.~\ref{fig:scheme} and~\ref{fig:ODMR_broadbandaddresing}(b), we probe the NV centers within a diamond sample with a broadband MW field. For the sake of simplicity, we consider the shape of the input broadband MW field to be Gaussian:
\begin{align} \label{eq:input_MW}
E_0(\omega) = \frac{1}{\sigma\sqrt{2\pi}} e^{\frac{-(\omega - \omega_0)^2}{2\sigma^2}},
\end{align}
where the central frequency $\omega_{0} = 2.87~$GHz matches the zero-field splitting $D$. When the MW pulse passes through the diamond sample, the response of the NV center spins will be approximately linearly proportional to the perturbation and characterized by dispersion $\chi_{1}(\omega)$ and absorption $\chi_{2}(\omega)$. The absorptive component can be found from the Lorentzian-decorated eigenfrequencies of the Hamiltonian. Provided that the absorption spectrum does not change during the pulse, $\chi_1(\omega)$ can be determined from $\chi_{2}(\omega)$ by Kramers-Kronig relations, yielding the susceptibility 
\begin{equation} \label{eq:response_function}
\chi(\omega) = \chi_{1}(\omega) + i \chi_{2}(\omega).
\end{equation}
The output signal as a result of the response function for a diamond sample with length $z$ and refractive index $n$ at MW frequencies would be:
\begin{align} \label{eq:output_MW}
E(z, \omega) = E_0(\omega) e^{\frac{2\pi}{c}iz(n\omega + \omega_0 \frac{\chi(\omega)}{2})}
\end{align}
The absorption profile of the output signal in the frequency domain is shown in Fig.~\ref{fig:ODMR_broadbandaddresing} (top panel) for applied magnetic fields of 2, 5, and 15 (G) chosen along a fiducial direction ($30^\circ$ polar and $60^\circ$ azimuthal angles relative to the crystallographic axis of the 100-oriented diamond). The input signal uniquely responds to each applied magnetic field. As the applied magnetic field strength increases, the absorption lines corresponding to each energy level spread out in the frequency domain. The top plot depicts a typical result from an ODMR measurement, where peaks correspond to absorption lines, while the bottom plot depicts the results from our scheme, where now troughs correspond to absorption. Even with a Gaussian envelope, the NV center absorption frequencies remain clearly visible. In our simulations, we used a Gaussian pulse with a temporal FWHM duration of $\sim$ 5 ns (corresponding to an 85 MHz FWHM bandwidth Gaussian pulse), which is sufficient to cover the full ODMR spectrum. Thus, the broadband illumination method can reproduce an ODMR measurement without requiring a frequency sweep.

For this method to work, at least two electronic spin states, either from the same or different NV orientations, must remain addressable by the MW field. This requires that the projection of the magnetic field onto different orientations does not result in frequency shifts that exceed ~100 MHz. The largest differences in the magnetic field projections may arise from the situation where the magnetic field is aligned with one of the NV orientations. In this case, with the electron gyromagnetic ratio of 28~GHz/T, one may detect external magnetic fields of up to 5.3~mT.

The microwave absorption coefficient for a waveguide attached to a diamond sample containing NV centers is computed in Appendix~\ref{app:MW abs coeff}. It provides the imaginary part of the magnetic susceptibility $\chi_2(\omega) \approx \chi_0\omega T_2^*/2$, under the conditions of resonance and sufficiently low MW drive power that power broadening is insignificant. Using the static susceptibility $\chi_0$, drive frequency $\omega$, and inhomogeneous dephasing time $T_2^*=1~\mu s$ appropriate to NV centers and realistic waveguide volumes and NV center concentrations of 4.5 ppm, the absorption is anticipated to be 1.28 dB/cm, with roughly 20\% of the incident MW power being absorbed per cm of propagation. We use a total absorption of 16\% in our simulations to demonstrate the results.

\section{Detection method and results}\label{results}
In practice, a fast detector reads the output signal in the time domain. Here, we use the Fourier transform to calculate the signal in the time domain, $E_{z}(t)$, from the result of Eq.~\ref{eq:output_MW}:
\begin{align} \label{eq:temporal_output_MW}
E(z,t) = \frac{1}{2\pi} \int_{-\infty}^\infty E(z,\omega) e^{-i \omega t} \, d\omega .
\end{align}
 \begin{figure}[htbp]
    \centering
    \includegraphics[width=1\columnwidth, viewport={7 2 547 420}, clip] {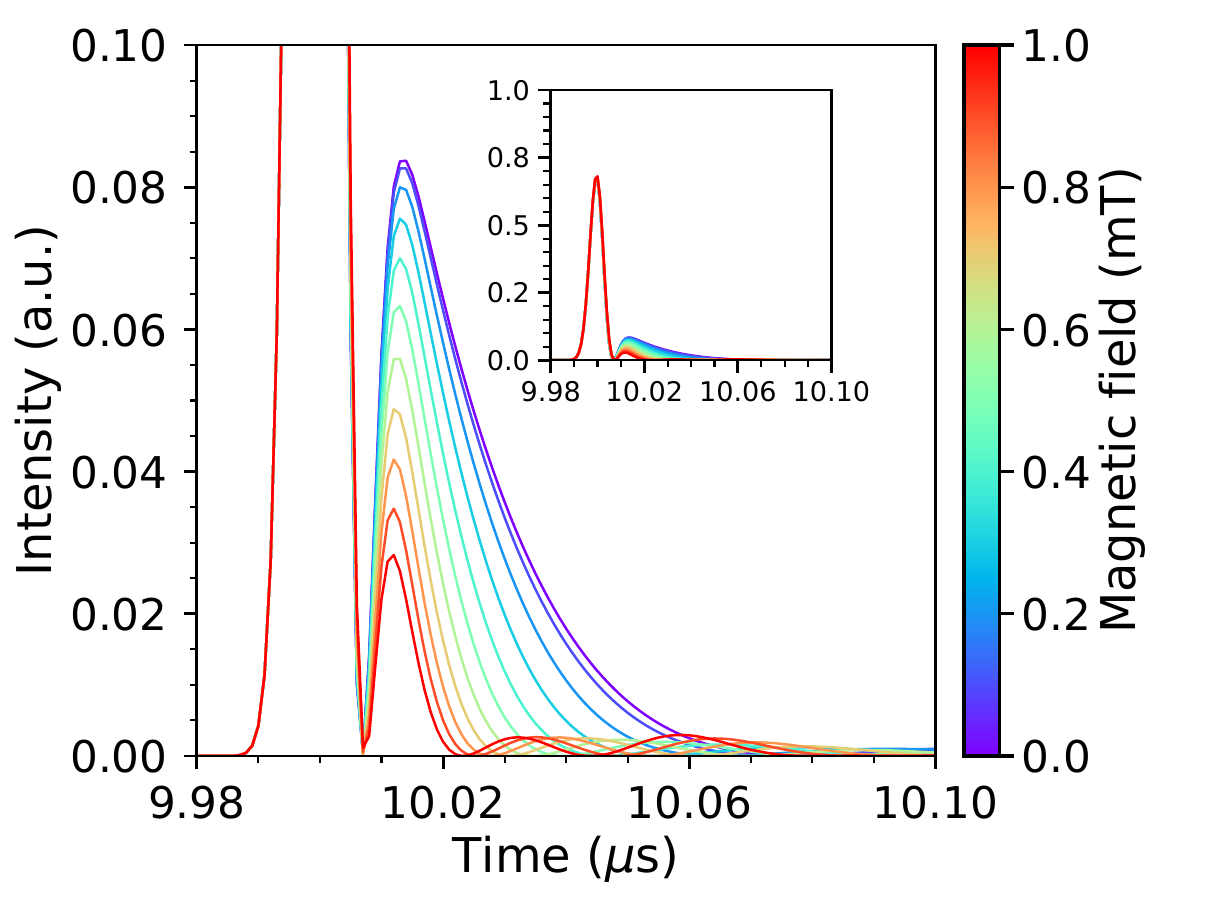}
    \caption{\justifying Time-domain MW transmission through an NV ensemble, $P(t)$, calculated using the same parameters as in Fig.~\ref{fig:ODMR_broadbandaddresing}, for applied magnetic fields ranging from 0 to 1~mT. The main panel shows a zoomed-in view of the signal to highlight the field-dependent variations, while the inset shows the same signal with the full intensity range (full $y$-axis). The color map indicates the applied magnetic field strength. Each field value produces a distinct transmission profile, illustrating how the NV ensemble response evolves as the external magnetic field changes.}
    \label{fig:signal_time_domain}
\end{figure}

Figure \ref{fig:signal_time_domain} shows the signal $P(t)\propto |E(z,t)|^2$ in the time domain 
for various external magnetic fields $B$. The signals exhibit different characteristics due to the changes in the response function that occur as the external magnetic field varies. The intensity information alone is sufficient to determine the magnetic field; phase information could be considered in a future study.

\subsection{Noise}
To investigate the sensitivity of our proposed approach, we add Gaussian random noise to the output signal. Later in this section and with further discussions on sensitivity in~Sec.~\ref{sensitivity}, we show how the signal-to-noise ratio (SNR) affects the absolute error in the predicted magnetic field, where the SNR in dB is given by $10 \log(\frac{\text{signal}_\text{power}}{\text{noise}_\text{power}})$.

In theory, we can input any amount of noise into our model. To make it realistic, we compute in Appendix~\ref{app:JN noise} the predicted amount of noise in our modeled system, following Ref.~\cite{eisenach2021cavity-enhanced}. Johnson-Nyquist noise far exceeds all other fundamental noise sources, such as spin-projection or MW shot noise~\cite{eisenach2021cavity-enhanced}, so we use the predicted Johnson-Nyquist noise to choose the SNR values in our computations. Our estimate of 75.5 dB for a signal of 50 mV indicates that high SNR values should be attainable in the laboratory. 
Noise can be further reduced by averaging the output signals over repeated measurements. These considerations are incorporated into our estimates of the total measurement time and overall sensitivity. The number of repetitions $R$ required to reduce the noise level of a signal from an initial standard deviation $\sigma_{\mathrm{SNR}_1}$ to a lower standard deviation $\sigma_{\mathrm{SNR}_2}$ is given by
\[
R = \left( \frac{\sigma_{\mathrm{SNR}_1}}{\sigma_{\mathrm{SNR}_2}} \right)^2,
\]
assuming uncorrelated Gaussian noise. Equivalently, expressing the improvement in terms of SNR values defined in decibels, the required number of repetitions is
\[
R = 10^{(\mathrm{SNR}_2 - \mathrm{SNR}_1)/10}.
\]

\subsection{Prediction models}
We employ two approaches to determine the magnetic field strength applied to the NV ensemble from the time-domain transmitted MW signal. First, we describe an approach based on the Kullback–Leibler (KL) divergence, where one can compare a measured output signal with a reference to find the best estimate of the underlying magnetic field. The second approach is based on neural networks, where we train our model with many examples of output signals for various magnetic field values. 

As seen below, we normalize the output intensity and treat it as a photon-arrival-time probability distribution to benefit from the KL divergence formalism. Though we expect to deal with high intensity MW fields, this approach is justified as even under noisy conditions we average over repeated experiments to reach a low-noise approximation of the signal before proceeding with the KL divergence approach. On the other hand, the estimation approach based on neural networks does not rely on any underlying assumptions about the signal that is fed into the model. Both of these approaches reach practical levels of sensitivity, with some neural network models performing better than others, all of which we will discuss below.

\subsubsection{Kullback–Leibler divergence}
The KL divergence, also known as relative entropy, quantifies the information loss incurred when using an estimated probability distribution \( Q(x) \) to approximate the observed data probability distribution \( P(x) \) within the same sample space \( X \), where $x\in X$ are the possible measurement outcomes. 
The KL divergence is mathematically defined as follows:
\begin{equation}\label{eq:KL_divergence}
D_{\text{KL}}(P \| Q) = \sum_{x \in X} P(x) \log\left(\frac{P(x)}{Q(x)}\right).
\end{equation}
In our simulation, the distributions \( P \) and \( Q \) are defined as:
\begin{align}\label{eq:probability_distribution}
P(t) = \frac{I(t)}{\sum_{t} I(t)}\quad \text{and} \quad Q(t|B) = \frac{J(t|B)}{\sum_{t} J(t|B)}.
\end{align}
Here, \( I(t) \)$=|E(z,t)|^2$ represents the observed temporal distribution, and \( J(t|B) \) is the expected temporal distribution of the output signal for a given magnetic field \( B \). Since, in the present study, even $I(t)$ comes from numerical simulations, both distributions are calculated in the time domain using Eqs. \ref{eq:output_MW} and \ref{eq:temporal_output_MW}, but in a physical experiment, $I(t)$ would be measured directly. 
This is why in our simulations we add noise to $I(t)$ but not to $J(t|B)$. 
In the calculation of KL divergence, the magnetic field \( B \) is varied for \( Q \), while it remains fixed for \( P(t) \), and a lower value of \( D_{\text{KL}} \) between the two distributions indicates that they are more similar.
By minimizing the divergence, we arrive at an estimator for the magnetic field:
\begin{equation}\label{eq:min_KL_divergence}
\tilde{B} = \arg\min_{{B_{model}}}D_{KL}[P(t) \| Q(t|B_{model})].
\end{equation} 
Since the distributions $P$ and $Q$ are interpreted as the likelihoods with which photons arrive at each time step, the value of $\tilde{B}$ is equivalent to the maximum-likelihood estimator for the magnetic field strength.

\begin{figure}[htbp]
    \centering
    \includegraphics[width=1\columnwidth, viewport={25 8 550 445}, clip] {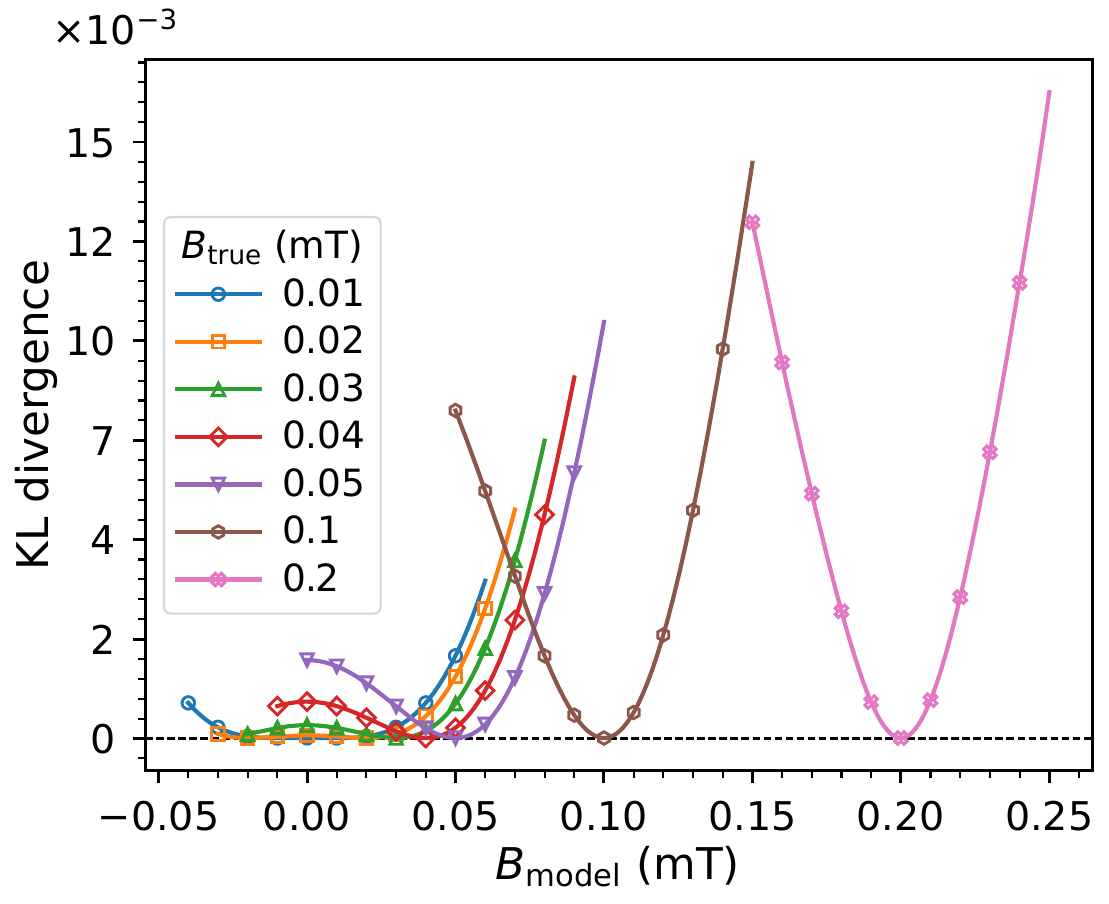}
    \caption{\justifying The KL divergence ($D_{KL}$) is computed across a range of magnetic field values applied to the probability distribution $Q(t| B_{model})$ while keeping the probability distribution $P(t)$ constant. The resulting curve reaches its minimum at the magnetic field strength where $P(t)$ and $Q(t| B_{model})$ exhibit the highest similarity. It should be noted that the granularity of the magnetic field, $B_{model}$, in the analysis is set at 
    10$^{-5}$~G; all other parameters are the same as in Fig.~\ref{fig:ODMR_broadbandaddresing}.}
    \label{fig:minimization_KLdivergence}
\end{figure}

Figure~\ref{fig:minimization_KLdivergence} illustrates the behavior of the KL divergence (\(D_{KL}\)) as a function of the modeled magnetic field \(B_{model}\), which as aforementioned, only leads to variation in the probability distribution \(Q(t| B_{model})\). In contrast, the reference probability distribution \(P(t)\) remained fixed because it depends on a fixed magnetic field $B_{true}$ with a single instance of small noise (SNR 100 dB) added to the intensity trace. The plot shows the KL divergence values for different values of \(B_{true}\), with each curve representing a distinct magnetic field condition.

As depicted, the KL divergence reaches its minimum at specific magnetic field strengths, indicating the optimal similarity between the reference distribution \(P(t)\) and the predicted distribution \(Q(t | B_{model})\). These minima indicate the points at which the model distribution best approximates the true underlying distribution and the value of $B_{model}$ at that minimum becomes the estimate $\tilde{B}$, with the narrowness of the curves and their unique minima highlighting the effectiveness of using KL divergence to identify the most representative magnetic field. Furthermore, the analysis was performed with a high level of precision, with the granularity of the predicted magnetic field (\(B_{model}\)) set at \(10^{-5}\). This fine granularity ensures that the computed KL divergence accurately captures variations across the range of magnetic fields tested, allowing for precise identification of the optimal similarity points.

It is also worth mentioning that both distributions $P$ and $Q$ are considered to have collected the signal over a $\Delta t=4\,\mu s$ time window. We tried different values of $\Delta t$ and found no marked difference in sensitivity per unit time for $\Delta t$ down to $1\,\mu s$, so long as the time around $10~\mu s$ after the incident pulse (see Fig~\ref{fig:signal_time_domain}) was included in the window. Then, for choosing the optimal start to this time window, we use a measure inspired by the Fisher information as explained in Appendix~\ref{app:fisher} to decide as to how long after the incident pulse one should measure the output signal in order to gain the most information about the underlying magnetic field.
The results show that there is a 1~$\mu s$ time window where the output intensity is most sensitive to changes in the external magnetic field so, to be safe, we measure between $8$~ and $12~\mu s$. This informs our subsequent choices about the measurement time per trial and eventually the total measurement time used to calculate the overall sensitivity of this approach.

\subsubsection{Neural network}
We also investigated the use of deep learning neural network models to estimate the value of $B$ from the time traces. We generated a dataset of 20000 time traces with SNRs of 20, 50, or 100 dB. 80\% of the dataset was used for training, 10\% for validation, and 10\% for testing. We examined several network architectures to perform this regression task, including:
\begin{enumerate}
    \item Multilayer perceptrons (MLP) with skip connections,
    \item 1D convolutional networks,
    \item Long short-term memory (LSTM) layers,
    \item Gated recurrent unit (GRU) layers,
    \item A latent neural ordinary differential equation (NODE) model.
\end{enumerate}

Details about model training (hyperparameter tuning, optimizers, learning rates and schedulers) are given in Appendix~\ref{app:NN model}. All models (besides the MLP) made use of either a single linear layer or an MLP to project the hidden state of the LSTM, GRU, or NODE layers to a single output, corresponding to the predicted magnetic field, akin to $\tilde{B}$ but using the deep-learning as an estimator. Notably, all variations of our convolutional models failed to learn from the data, and consistently predicted the mean of the dataset labels (i.e. a value of 0.5 for the normalized $B$ range of 0.0 to 1.0). We observed this for other hybrid models where any convolutional layers were present, e.g. downsampling with convolutional layers before or after recurrent layers, implying that these steps were degrading or removing useful information about $B$.

The best performance is attained using the NODE model, followed by the GRU, LSTM, and MLP models, respectively. None of the performances are as good as the KL divergence method; for the SNRs that we simulated, the sensitivities of the best NODE model are within a factor of 10 of the KL method. This is sensible because the KL method corresponds to maximum likelihood estimation and should provide an optimal unbiased estimator in the large-sample limit. Nonetheless, machine learning methods should shine in situations where sample data is limited, where KL minimization is computationally expensive, or where empirical calibration is needed for capturing all of the noise in dynamics. 


\section{Sensitivity}\label{sensitivity}
The sensitivity of NV centers in diamond to magnetic fields is a critical element of magnetometry. This sensitivity informs the minimum magnetic field that an NV ensemble can detect within a fixed time for a given approach. Typically, sensitivity is influenced by factors such as coherence time (here we take $T_2^*=1~\mu s$) and photon collection efficiency. Our method, which does not rely on optical readout, complicates conventional sensitivity calculations due to factors such as SNR levels that depend on the intensity of the MW input, density of the NV centers to avoid saturated absorption, and MW waveguide or cavity design and characteristics.
 \begin{figure}[htbp]
    \centering
    \includegraphics[width=1\columnwidth, viewport={7 8 550 445}, clip]{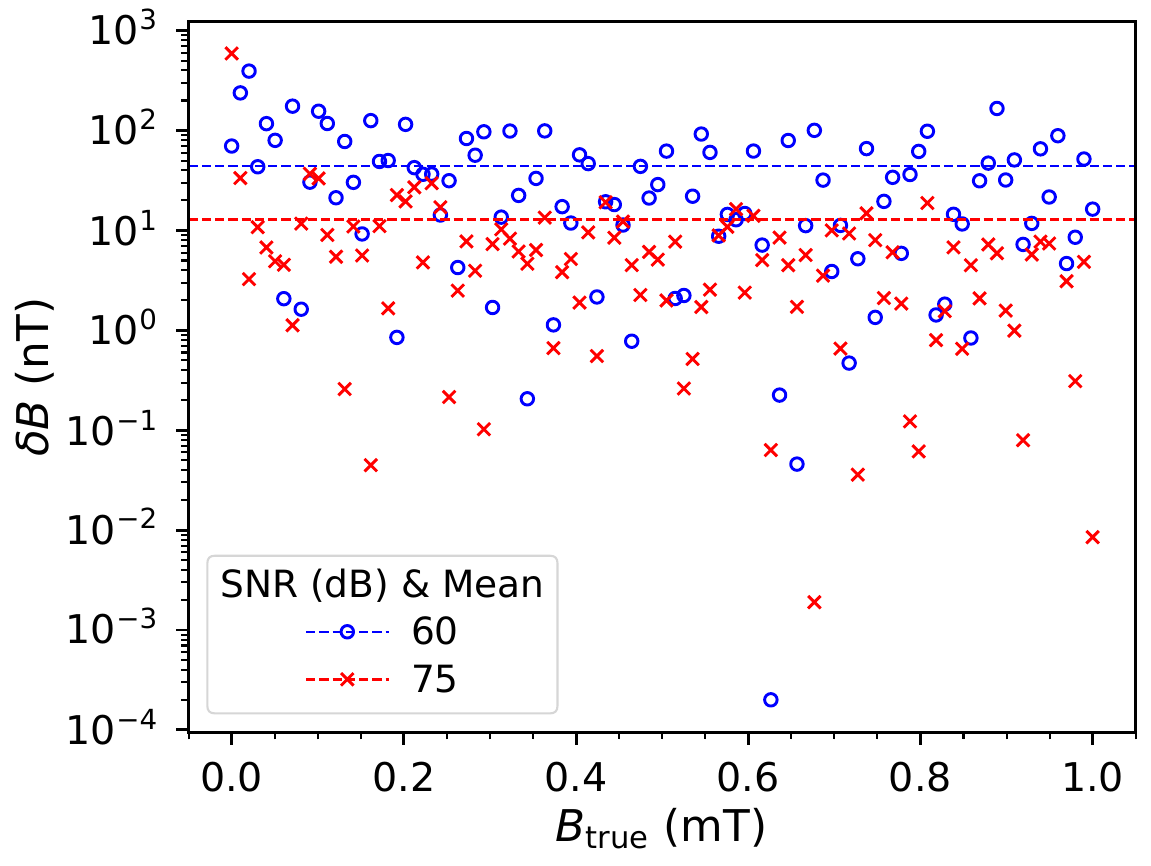}
    \caption{\justifying Absolute magnetic field estimation error, \(\delta B = |\tilde{B} - B_{\text{true}}|\) , as a function of the true magnetic field $B_{\text{true}}$ for various signal-to-noise ratios (SNR) in dB. The vertical axis is displayed on a logarithmic scale to highlight variations across several orders of magnitude. Scatter markers represent individual data points for each SNR, while the dashed horizontal lines indicate the mean $\delta$B for the corresponding SNR. The results show a strong decrease in $\delta$B with increasing SNR.}
    \label{fig:Boptimal_vs_Breal_various_SNR}
\end{figure}
Here, we study magnetic field sensitivity, as well as its accuracy, by evaluating the quality of the magnetic field extraction using KL divergence calculations for a signal with low noise (high SNR). 

The ultimate bound on the sensitivity achievable with this method is given by the curvature of the KL divergences versus true magnetic field and the number of photons per second. Those curvatures are on the orders of $10^{-3}$ to $10^{-1}~\mathrm{G}^{-2}$ for true magnetic fields ranging from 0.005 to 0.2~mT which, with our rates of $5\times 10^{13}$ photons per $5\,\mu s$ pulse, could achieve shot-noise-limited sensitivities ranging from 10 $\mathrm{fT/\sqrt{Hz}}$ to 1 $\mathrm{pT/\sqrt{Hz}}$. A real experiment will be limited by factors other than this shot noise and so we study the effects of realistic SNR values stemming from Johnson-Nyquist noise.

Beginning with an example, we compare in Fig.~\ref{fig:Boptimal_vs_Breal_various_SNR} the magnetic field predicted by the KL divergence, denoted as $\tilde{B}$, and the actual magnetic field, \(B_{\text{true}}\), for different SNRs. SNR is quantified as the ratio of the maximum signal of the output to the Gaussian noise width. The accuracy of the calculated prediction \(\tilde{B}\) improves with higher SNR values. 

\begin{figure}[htbp]
    \centering
    \includegraphics[width=\columnwidth, viewport=7 5 580 447, clip]{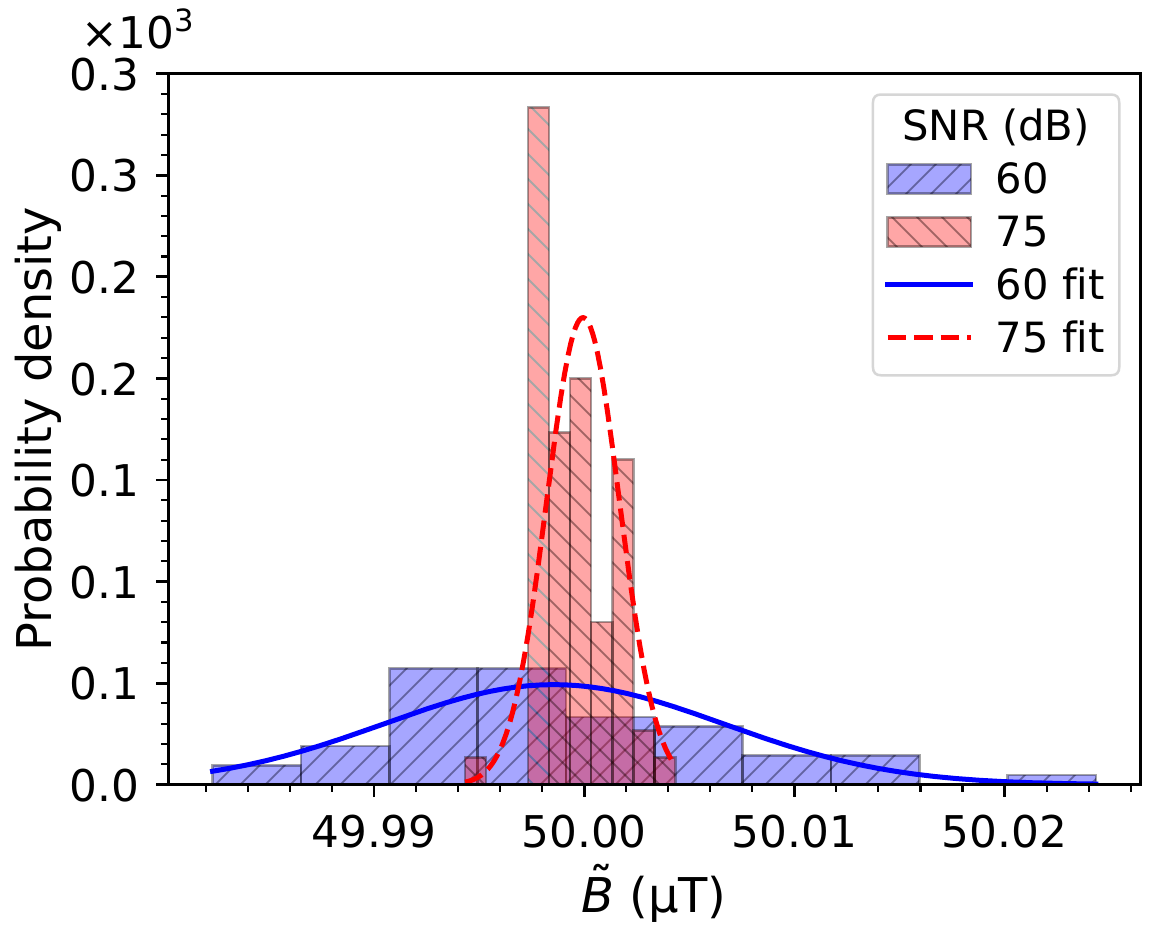}
    \caption{\justifying Histogram of $\tilde{B}$ values with Gaussian fits for an ensemble of NV centers, measured under a real magnetic field of 50~$\mu$T (0.05~mT) at SNRs of 60~dB and 75~dB. Each histogram reflects the distribution and precision of $\tilde{B}$ estimations at increasing levels of signal clarity.}
    \label{fig:normal_distribution_Bopt_various_SNR}
\end{figure}

Figure~\ref{fig:Boptimal_vs_Breal_various_SNR} concerns a single instance of the random noise for each SNR value. To evaluate the overall sensitivity and accuracy of our method, we repeat this process for 50 instances of the random noise and observe that the calculated predictions are consistently close to the true values.
For SNR values of 60 and 75~dB, we used the KL-minimization technique to predict the magnetic field value when the true value was 0.5 G, then plotted histograms of the predictions in
Fig.~\ref{fig:normal_distribution_Bopt_various_SNR}. As the SNR increases from 60 to 75~dB, the histograms narrow.
The uncertainties in the estimated magnetic fields, denoted \(\sigma_B\), 
correspond to the standard deviations of Gaussian distributions fitted to the histograms of predicted magnetic field values and are also illustrated in Fig.~\ref{fig:normal_distribution_Bopt_various_SNR}.

The values of the mean predicted magnetic field \(\tilde{B}_{\text{mean}}\) and uncertainty \(\sigma_B\) at an SNR of 75~dB are presented in Table~\ref{tab:error_B0_Bopt}. This analysis covers a range of true magnetic fields from 0.01~mT to 0.1~mT. The KL divergence sensitivity is computed using 100{,}000 data points sampled across this magnetic field range. For these parameters, the accuracy and sensitivity are characterized by biases 
$\Delta B = |B_{\mathrm{true}} - \tilde{B}_{\mathrm{mean}}|$ 
in the range of $0.03$–$0.7~\mathrm{nT}$ and uncertainties 
$\sigma_B$ in the range of $1.4$–$4~\mathrm{nT}$.

\begin{table}[htbp]
    \centering
    \caption{\justifying Comparison of the true magnetic field (\(B_{\text{true}}\)) and the average estimated magnetic field (\(\tilde{B}_{\text{mean}}\), obtained by minimizing the Kullback–Leibler (KL) divergence for randomized SNR = 75~dB, across various magnetic field strengths in the range of 0.01 to 0.1~mT. The average bias \(\Delta B = |B_{\text{true}} - \tilde{B}_{\text{mean}}|\) quantifies the accuracy of the estimation and is around $10^{-5}$ G for each true field value. The uncertainty \(\sigma_ B\) represents the standard deviation in the estimated magnetic field after averaging over multiple measurements and is calculated from a Gaussian fit to the predicted field distribution. The sensitivity is computed for a single measurement with 100~$\mu s$ reset time.}

\begin{tabular}{c c c c}
        \toprule
        \(B_{\text{true}}\) (mT) & \(\Delta B\) (nT) & \(\sigma_B\) (nT) & 
        \begin{tabular}[c]{@{}c@{}} \(\sigma_B \times \sqrt{\text{time}}\) (pT/\(\sqrt{\text{Hz}}\))

        \\ 
        \end{tabular}
        
        \\

        0.01 & 0.7 & 3.9(4) & 39 \\
        0.02 & 0.2 & 2.2(2) & 22 \\
        0.03 & 0.5 & 1.6(2) & 16 \\
        0.04 & 0.2 & 1.4(1) & 14 \\
        0.05 & 0.1 & 1.4(1) & 14 \\
        0.06 & 0.5 & 1.4(1) & 14 \\
        0.07 & 0.6 & 1.5(2) & 15 \\
        0.08 & 0.6 & 1.6(2) & 16 \\
        0.09 & 0.03 & 1.6(2) & 16 \\
        0.1 & 0.2 & 1.5(2) & 15 \\
        
    \end{tabular}
    \label{tab:error_B0_Bopt}
\end{table}
In addition, we analyze the total time required to achieve an SNR of 75 dB in a single-shot measurement. In this case, the experimental cycle is dominated by a 100 $\mu s$ reset time, and no averaging over multiple measurements is performed. The corresponding results are also summarized in Table~\ref{tab:error_B0_Bopt}, expressed as precision per unit time.

These inform a fully realistic evaluation of our method. Using measured properties of diamond samples and their interactions with waveguide-coupled light outlined in Appendix~\ref{app:JN noise}, the dominant fundamental source of noise is Johnson-Nyquist noise of the detectors, with effects such as shot noise being relatively negligible. Over the course of a single experimental duty cycle, Johnson-Nyquist noise can lead to an SNR of approximately 75~dB (as discussed in \ref{app:JN noise}), a process that can be repeated until the SNR improves statistically to 100~dB after 158~$\mu s$. Despite this, we limit our results in Table~\ref{tab:error_B0_Bopt} to SNR of 75~dB that is more in line with MW source noise and limitations from sampling of the output signal; see below for a discussion on avenues for improvement.

The KL method thus provides achievable sensitivities on the order of $10~\mathrm{pT/\sqrt{Hz}}$ and achievable accuracies well below $1~\mathrm{nT}$ (i.e., we see systematic errors below $1~\mathrm{nT}$ associated with the KL minimization routine) for measuring magnetic fields ranging from 0.01 to 0.1~mT with update rates up to 10~kHz, with further improvement possible by refining the pulse shape and other operational parameters. In the absence of a complete model that would capture the effects of strain and temperature, one can rely on experimental calibration to establish the reference signals needed for the KL method.

We limited our sensitivity calculations above to an SNR of 75~dB. 
While we estimate an SNR of 75.5~dB from Johnson-Nyquist noise for a 50~mV signal, this MW intensity is orders of magnitude below saturation and line-broadening limits and could be increased, thus reducing the impact of Johnson-Nyquist noise.  In practice, the SNR will likely be limited by technical noise: for example, current digitizers offer $\sim$~70~dB SNR at 1~ns sampling (see Appendix~\ref{app:JN noise}). Future refinements in MW instrumentation along with higher MW signal intensities could thus significantly improve achievable SNR and predicted sensitivities. 
This is especially true because the shot noise remains below 1 $\mathrm{pT/\sqrt{Hz}}$. 

In addition, optimizing pulse shaping and other operational parameters would allow further improvements in the sensitivity. With these considerations, sensitivities down to 10~pT/$\sqrt{\text{Hz}}$ may become achievable with our proposed method.
We note that while our method performs well at low magnetic fields (down to approximately 0.1 G in our simulations), it does not extend to strictly zero magnetic field. At zero bias field, the degeneracy of the $m_s = \pm 1$ states leads to vanishing first-order magnetic sensitivity, analogous to conventional ODMR. In realistic diamond samples, transverse strain or electric fields further mix these states, resulting in a characteristic scale \(B \lesssim d_\perp E_\perp / (g\mu_B)\) below which magnetic sensitivity degrades. For typical parameters, this corresponds to bias fields on the order of a few tenths of a Gauss or less. Our simulations therefore focus on the experimentally relevant regime of low but finite magnetic fields, where magnetic sensitivity remains robust.

\section{Conclusion}
We proposed a new approach to probe NV centers in diamond for magnetic field sensing using broadband MW pulses. We demonstrated that the time-domain transmitted MW field does carry information about the absorption and dispersion due the electronic and hyperfine states of NV centers which, consequently, allow us to develop a measure for the external magnetic field affecting these states. We utilized the minimization of the KL divergence to obtain a maximum likelihood estimate of the underlying magnetic field, yielding sensitivities that are on the order of 10~pT/$\sqrt {Hz}$; down to the $\sim$1~nT levels we investigated, the analysis introduced no detectable systematic error. Our approach can potentially be further improved by optimizing the MW probe pulse shape and bandwidth. We expect our approach to be extendable to vectorial magnetic detection. One key aspect of our proposed approach to highlight is that, unlike ODMR, we do not require a bias magnetic field, as our approach remains reliable down to magnetic fields of 0.01~mT (or where strain-induced effects become dominant). 
We believe our broadband addressing proposal that simultaneously eliminates the optical detection requirement will help adoption of this approach for practical devices.
\begin{acknowledgments}
    The NRC headquarters is located on the traditional unceded territory of the Algonquin Anishinaabe and Mohawk people. AA, LC, and KH thank Romain Ruhlmann, Vincent Halde, Yves B\'erub\'e-Lauzi\`ere, and Ankita Chakravarty for helpful discussions and comments. KH acknowledges funding from the NSERC Discovery Grant and Alliance programs. LC acknowledges funding from the NSERC Discovery Grant program. This work was supported by NRC's Quantum Sensors challenge program.
\end{acknowledgments}

%

\appendix

\renewcommand{\thefigure}{S\arabic{figure}}
\setcounter{figure}{0}

\section{Supplementary Materials}
\subsection{Microwave Absorption Coefficient}
\label{app:MW abs coeff}

To estimate the microwave (MW) absorption by NV$^-$ centers in diamond at the resonance frequency corresponding to spin transitions (in the absence of an external magnetic field), we follow the approach outlined in \cite{hagen2013broadband}. The MW absorption coefficient is given by:

\begin{equation}\label{eq:MW_absorption}
\alpha \simeq 27.3 \cdot \frac{\sqrt{\epsilon_R}}{\lambda_0} \cdot \chi_2,
\end{equation}  
where \(\epsilon_R \simeq 5.7\) is the relative permittivity of diamond at microwave/EPR frequencies, \(\lambda_0\) is the free-space wavelength of the MW field, and \(\chi_2\) is the imaginary part of the magnetic susceptibility.

The imaginary part of the susceptibility for a two-level spin system, such as the NV center, is given by \cite{eisenach2021cavity-enhanced}:

\begin{equation}\label{eq:imaginary_susceptibility}
\chi_2(\omega_d) = \frac{1}{2} \chi_0 \cdot \frac{\omega_s T_2^*}{1 + (\omega_d - \omega_s)^2 T_2^{*2} + \left( \frac{\gamma B_1}{2} \right)^2 T_1 T_2^*},
\end{equation}
where \(\omega_d\) is the drive angular frequency and \(\omega_s\) is the spin resonance frequency. Here, \(\chi_0\) is the static susceptibility, \(T_2^*\) is the transverse relaxation time (spin dephasing), \(T_1\) is the longitudinal relaxation time (spin-lattice relaxation), \(\gamma\) is the gyromagnetic ratio, and \(B_1\) is the amplitude of the applied MW magnetic field. The denominator includes contributions from detuning \((\omega_d - \omega_s)\) and power broadening due to the MW drive field \(B_1\), reflecting how susceptibility varies with frequency and MW power.

Under the on-resonance condition \((\omega_d = \omega_s)\) and assuming low MW drive power such that \(\gamma B_1 \approx 0\), equation \eqref{eq:imaginary_susceptibility} simplifies to:

\begin{equation}\label{eq:reduced_imaginary_susceptibility}
\chi_2(\omega_s) \approx \frac{1}{2} \chi_0 \omega_s T_2^*.
\end{equation}

The static magnetic susceptibility \(\chi_0\) describes the linear response of a spin ensemble to a weak static magnetic field and is given by:

\begin{equation}\label{eq:static_susceptibility}
\chi_0 = \frac{\mu_0 (g \mu_B)^2}{\hbar \omega_s} \cdot \frac{N_0 - N_{+1}}{V},
\end{equation}
where \(\mu_0\) is the vacuum permeability, \(g\) is the Landé g-factor, \(\mu_B\) is the Bohr magneton, and \(V\) is the volume of the diamond. The terms \(N_0\) and \(N_{+1}\) represent the spin populations in the \(|m_s = 0\rangle\) and \(|m_s = +1\rangle\) states, respectively. Their difference corresponds to the number of polarized spins contributing to MW absorption. Although the volume \(V\) appears in the expression for static susceptibility (equation \eqref{eq:static_susceptibility}), it cancels out when the spin polarization is expressed in terms of a volumetric density (e.g., spins per $m^3$). In our calculation, we directly use the polarized spin density, which already accounts for the NV concentration, charge state occupancy, spin polarization fraction, and the intrinsic atomic density of diamond. However, the physical dimensions of the waveguide
are important as they define the mode volume and the total number of accessible NVs, which would set the limit on the signal intensity based on the saturation threshold. A range of waveguide parameters would work with a mode volume of approximately 0.5~mm$^{3}$ (total number of NVs approximately $4.2\times 10^{14}$). 

In the ideal scenario, several favorable conditions are met to maximize MW absorption in NV-doped diamond. These include full polarization of all NV centers in the \(|m_s = 0\rangle\) state, 100\% occupancy of the negatively charged NV$^-$ state, MW driving in resonance \((\omega_d = \omega_s)\), and negligible power broadening \((\gamma B_1 \to 0)\). Furthermore, a high density of NVs and a long coherence time \(T_2^*\) enhance absorption, while optimal spatial overlap between the MW mode and the NV volume ensures efficient energy transfer. Under such ideal conditions, MW absorption can reach \(2~\text{dB/cm}\) per ppm of fully polarized NV$^-$ centers, calculated from the theory in \cite{eisenach2021cavity-enhanced}.

For comparison, a more realistic scenario involves commercially available diamonds such as the E6 B14, which is widely used in many magnetometry experiments. This diamond typically contains an NV concentration of approximately 4.5 ppm, or about 1 ppm per crystallographic orientation. In practice, only about 70\% of the NV centers are in the negatively charged NV$^-$ state, and of those, roughly 80\% are spin-polarized into the \(|m_s = 0\rangle\) state under optical pumping. When accounting for these realistic charge and spin polarization fractions, the effective polarized NV density is reduced to approximately 0.56 ppm. As a result, a more practical estimate for MW absorption is around \(1.34~\text{dB/cm}\), representing a reasonable estimate under typical experimental conditions. This level of absorption corresponds to approximately 27\% of the incident MW power that is absorbed per centimeter of propagation through the material.

\subsection{Absorption dependence of the transmitted microwave response}

\begin{figure*}[t]
    \centering

    \begin{minipage}{0.32\textwidth}
        \centering
        \includegraphics[width=\linewidth, viewport={8 2 552 500}, clip]
        {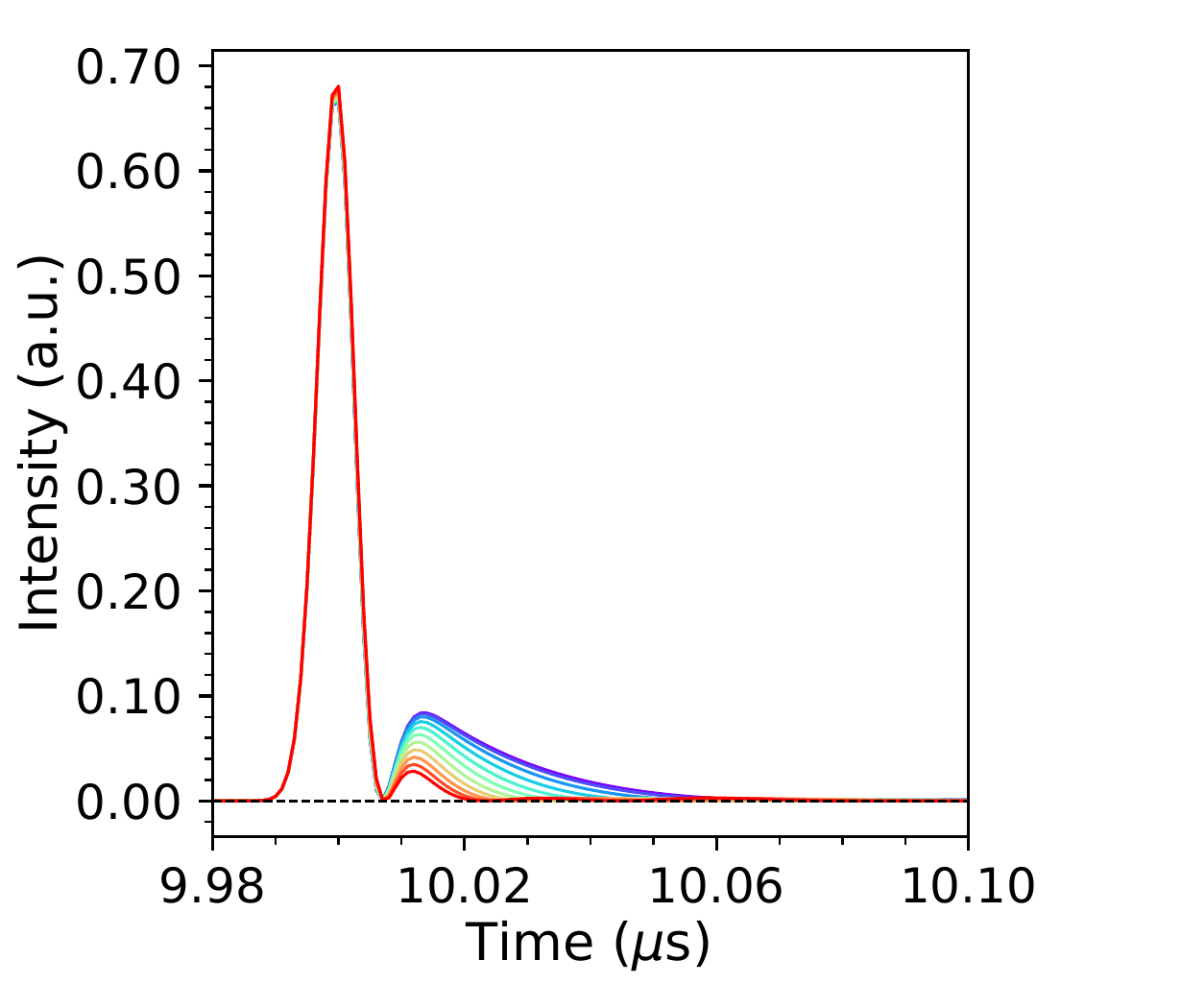}
    \end{minipage}
    \hfill
    \begin{minipage}{0.32\textwidth}
        \centering
        \includegraphics[width=\linewidth, viewport={8 2 552 500}, clip]
        {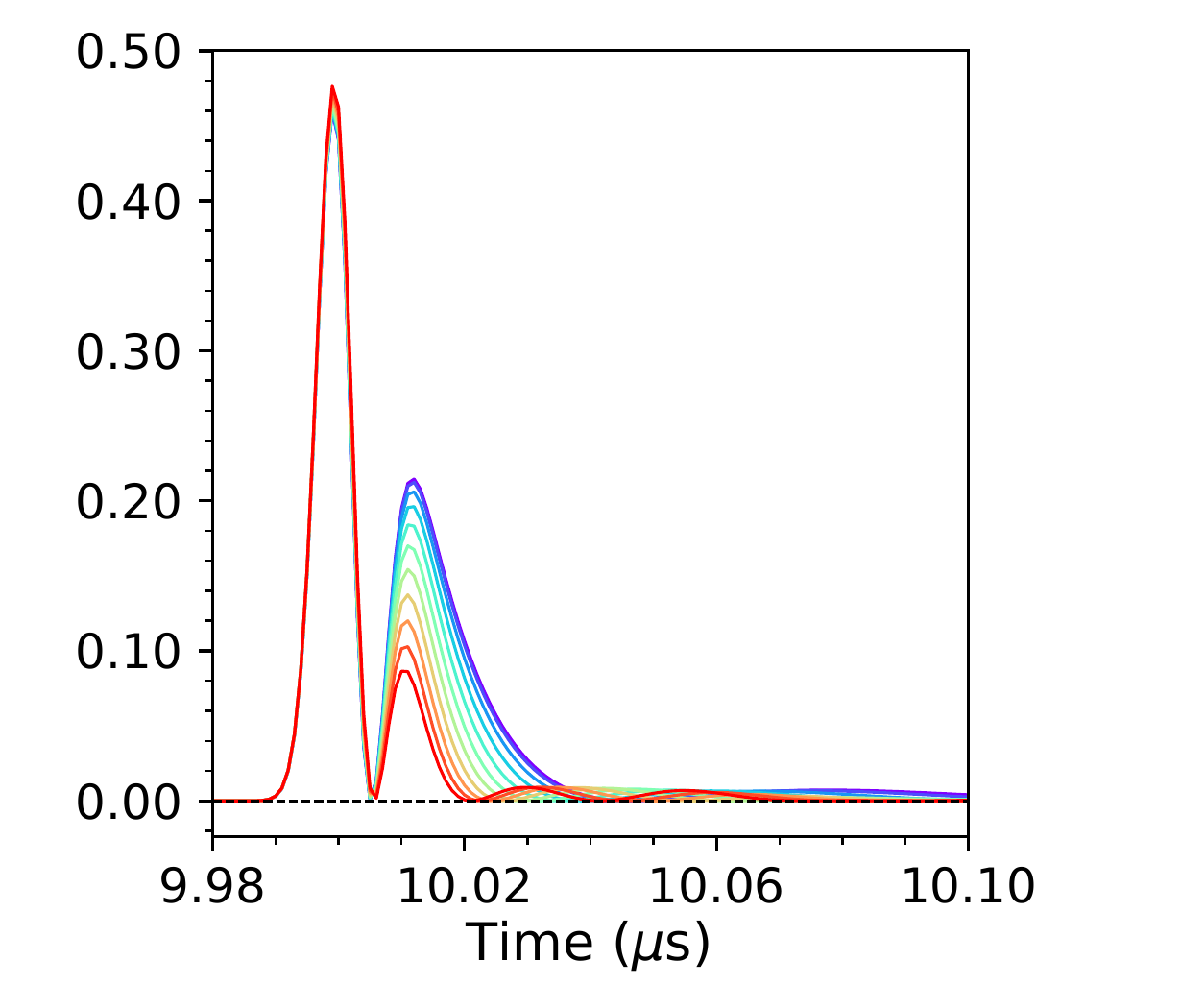}
    \end{minipage}
    \hfill
    \begin{minipage}{0.32\textwidth}
        \centering
        \includegraphics[width=\linewidth, viewport={8 2 552 500}, clip]
        {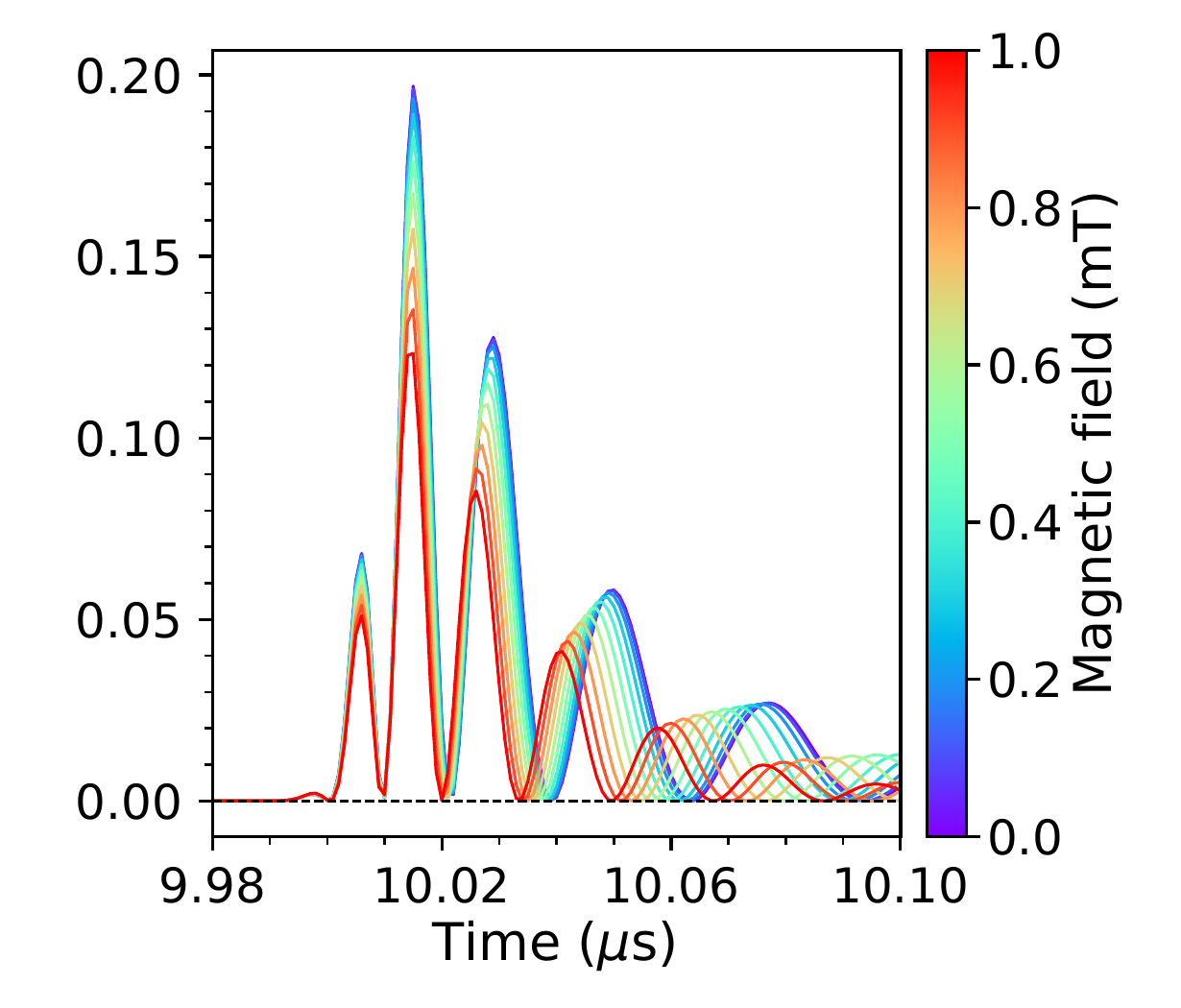}
    \end{minipage}

    \caption{\justifying
    Transmitted microwave intensity as a function of time for magnetic fields between 0 and 1~mT at different microwave absorption levels. The left, middle, and right panels correspond to approximately 16\%, 20\%, and 47\% absorption, respectively. Note the different ranges on the vertical axes, with higher peaks at lower absorptions.
    }
    \label{fig:absorption_time_trace}
\end{figure*}

To further clarify the field-dependent structure in the transmitted microwave signal, we compare the time-domain response for different microwave absorption values. Figure~\ref{fig:absorption_time_trace} shows the transmitted microwave intensity as a function of time for different magnetic field strengths. The left, middle, and right panels correspond to approximately 16\%, 20\%, and 47\% microwave absorption, respectively. As the absorption increases, more structure becomes visible in the transmitted signal. In particular, the field-dependent temporal reshaping and beating-like features become clearer at higher absorption. This indicates that stronger absorption enhances the spectral filtering imposed by the NV ensemble on the broadband microwave pulse.

This improvement in the visibility of the temporal features, however, comes with a practical trade-off. While higher absorption makes the field-dependent structure easier to distinguish, it also reduces the transmitted microwave signal strength. In the presence of detector noise, this reduction in signal amplitude can decrease the amount of useful information available in the measured waveform, even though the waveform appears more structured. Therefore, increasing the absorption does not necessarily lead to a monotonic improvement in field estimation. Instead, the optimal absorption depends on a balance between enhancing the field-dependent temporal contrast and maintaining enough transmitted signal strength for reliable detection.

\subsection{Optical Pumping Power Estimate}\label{app:optical_power_estimate}
Optical spin polarization is required to initialize the NV ensemble prior to each measurement cycle. We estimate $m \approx 3$ photons are required per NV center to achieve spin polarization into the $|m_s = 0\rangle$ state~\cite{barry2020sensitivity}. 

We consider a sensing volume corresponding to typical transverse waveguide dimensions and propagation length of $1~\text{cm} \times 500~\mu\text{m} \times 100~\mu\text{m}$. For an NV density of 4.5 ppm, the number of NV centers within this volume is approximately $N \approx 4.2 \times 10^{14}$.

Assuming optical pumping at $\lambda = 532~\text{nm}$, the total energy required to polarize the ensemble is
\begin{equation}
\label{eq:pumping_energy}
E = N m \frac{hc}{\lambda},
\end{equation}
which evaluates to approximately $E \approx 470~\mu\text{J}$.

If polarization is completed within the $100~\mu\text{s}$ reset time used throughout the manuscript, this corresponds to an absorbed optical power of approximately $4.7~\text{W}$.

\subsection{Detector and Source Requirements for Experimental Implementation} \label{app:implementation}
To assist the experimental implementation, we summarize the MW detector and source specification required to reproduce the theoretical performance discussed in the main text.

\subsubsection{Johnson-Nyquist Noise and Signal-to-Noise Considerations}
\label{app:JN noise}

Johnson-Nyquist noise, also known as thermal noise, arises from the random thermal motion of charge carriers (typically electrons) in a resistor. It is present in the microwave detector even in the absence of any applied voltage or current and represents a fundamental noise floor in electronic systems. The voltage generated on a resistor \(R\) over a bandwidth \(\Delta f\) at temperature \(T\) is given by the following:

\begin{equation} \label{eq:johnson_noise}
V_{\text{JN}} = \sqrt{4 k_B T R \Delta f}.
\end{equation}
where \(k_B\) is the Boltzmann constant. For a system at room temperature (\(T = 300~\text{K}\)), with a termination resistance of \(R = 50~\Omega\) as the MW couples to the detector and a field bandwidth of broadband MW of \(\Delta f = 85 \times 10^6~\text{Hz}\), the Johnson noise voltage is calculated as:

\[
V_{\text{JN}} \approx 8.4 ~\mu\text{V}.
\]

This value defines the thermal noise voltage floor of the system and typically dominates other noise sources~\cite{eisenach2021cavity-enhanced}.
To evaluate the impact of this noise on signal quality, the signal-to-noise ratio (SNR) can be expressed in decibels (dB) as:
\begin{equation} \label{eq:johnson_noise_SNR}
\mathrm{SNR}_{\text{dB}} = 10 \log_{10} \left( \frac{P_s}{P_n} \right),
\end{equation}
where \( P_s \) and \( P_n \) are the signal and noise powers, respectively, each calculated using \( P = V^2 / R \).
Assuming an applied microwave field with rms voltage  \( V_{\mathrm{rms}} = 0.05~\text{V} \), as reported in~\cite{eisenach2021cavity-enhanced}, the corresponding signal-to-noise ratio with respect to thermal noise is:
\[
\mathrm{SNR}_{\text{dB}} = 10 \log_{10} \left( \frac{(0.05)^2}{(8.4 \times 10^{-6})^2} \right) \approx 75.5~\text{dB}.
\]
This high SNR indicates that under these conditions, the signal is well above the thermal noise floor, enabling high-fidelity detection in the system.

\subsubsection{Temporal Resolution}
\label{app:temporal_resolution}
The simulated time-domain signals were generated at 1~ns temporal resolution. We averaged the data to simulate 10~ns sampling and re-evaluated the KL divergence at an applied magnetic field of 0.5 G with an SNR of 65~dB for each sample, such that the total 10~ns sample effectively has 75~dB SNR. Under these conditions, the resulting magnetic field uncertainty is approximately 4~nT. In comparison, performing the calculation directly at 1~ns temporal resolution with the same SNR yields a smaller uncertainty of 1.4~nT, which is close to $\sqrt{10}$ smaller. This indicates that the temporal features have a characteristic scale of at least 10~ns and thus that high-quality 10~ns sampling is sufficient to reproduce the primary conclusion of this work. 

Achieving $\sim$ 75~dB SNR with 1-10~ns sampling rates is near the upper end of current commercial RF digitizer capabilities. Presuming signals have been mixed down with a high-SNR pure tone, digitizers approaching this regime are available.
\footnote{If amplification is required to match the digitizer's full-scale input range, low-noise RF amplifiers are available. For example, a 19.5~dB gain amplifier with a 1.8 dB noise figure would degrade the system SNR by approximately 1.8~dB, provided it dominates the noise chain. With an anticipated signal level of $\sim$0.05~V peak ($\approx$ -16~dBm into 50~$\Omega$), approximately 20~dB of gain is sufficient to utilize the full input range of typical high-speed digitizers.}
For example, Spectrum Instrumentation reports 73.3~dB SNR at 125~MS/s and the Teledyne SP Devices ADQ1600RF provides 1.6~GS/s sampling, 750~MHz analog bandwidth, and approximately 69~dB SNR.

\subsubsection{Signal Generation Requirements} \label{app:signal_generation}
The broadband MW field is assumed to have an 85~MHz bandwidth in our simulations. In practice, a bandwidth on the order of 100~MHz is required to generate the probe pulse. High-performance arbitrary waveform generators (AWGs) capable of this bandwidth are commercially available. For example, the Zurich Instruments HDAWG (750~MHz analog bandwidth) specifies 0.32~mV$_{\mathrm{rms}}$ output voltage noise on a $\pm 2.5$~V range, corresponding to an SNR of approximately 84~dB relative to full-scale output.

In practice, the baseband waveform must be upconverted by mixing with a low-phase-noise microwave carrier. Commercial microwave sources such as QuSine report phase noise as low as $-154$~dBc/Hz. When integrated over a 100~MHz bandwidth, this phase-noise level corresponds to an effective carrier SNR on the order of $\sim 70$--$75$~dB, depending on the integration bandwidth and detailed noise spectrum. These specifications indicate that generation of broadband MW pulses with SNR in the 70--80~dB range is feasible using existing commercial hardware.

\subsection{The Classical Fisher Information}
\label{app:fisher}
Choosing an appropriate time window to estimate the magnetic field is crucial. A well-sized and positioned window extracts most of the information from the signal without unnecessarily constraining the repetition rate of the experiment and limits the computational resources needed to store and analyze the data. Figure~\ref{fig:signal_time_domain} in the main text shows that most of the variation in the signal occurs around $10~\mu \mathrm{s}$ after the applied magnetic field, which serves as the starting point for our analysis. 

We employ a heuristic based on Fisher information to explore a which times in the temporal signal contribute most to an estimate of the magnetic field.
The Fisher information $F(\theta)$ describes how helpful a measurement is to estimate small changes in an unknown parameter $\theta$. A measurement that yields an outcome $i$ with probability $p_i(\theta)$ has
\begin{equation}\label{eq:classical_Fisherinfo}
F(\theta) = \sum_{i} \frac{1}{p_{i}(\theta)} \left( \frac{\partial p_{i}(\theta)}{\partial \theta} \right)^2.
\end{equation}
Intuitively, the Fisher information sums up how strongly each outcome probability varies with small changes in $\theta$, with rare events receiving more weight.

\begin{figure}
    \centering
    \includegraphics[width=1\columnwidth, viewport={7 1 550 445}, clip]{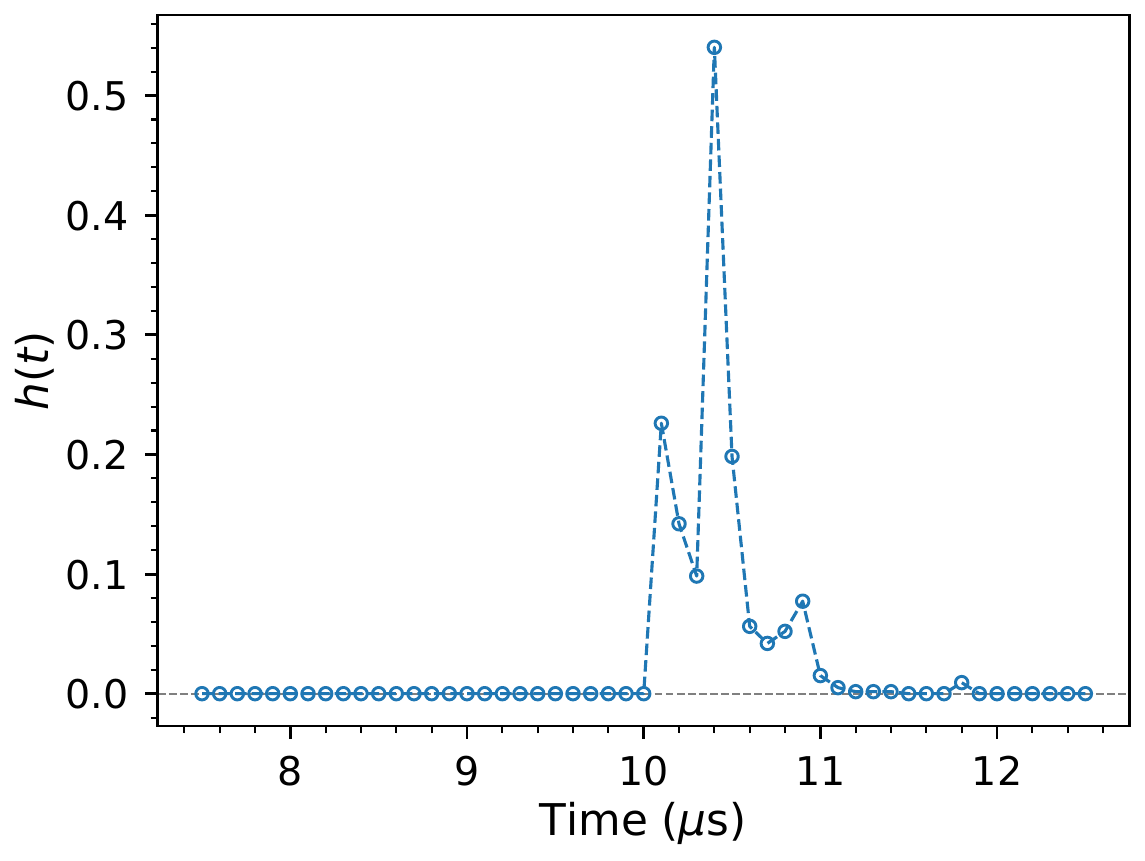}

    \caption{\justifying The function h(t) was calculated from time-domain signal profiles acquired while sweeping the magnetic field from 0 up to 5 G. For each field value, the signal intensity at specific time points was extracted, and h(t) was obtained by summing these intensities over all magnetic fields at each time point. In this construction, h(t) reflects how the temporal response evolves with applied magnetic field. It does not provide meaningful information for times shorter than about 10 µs or longer than about 11 $\mu$s; only the interval between 10–11 $\mu$s should be regarded as reliable.}
    \label{fig:CFI_broadbandaddresing}
\end{figure}

Inspired by Equation \ref{eq:classical_Fisherinfo}, we use the expression
\begin{equation}
    \frac{1}{J(t|B)} \left ( \frac{J(t|B + \Delta B) - J(t|B)}{\Delta B} \right )^2
\end{equation}
to quantify how helpful sampling the signal at time $t$ is for estimating small variations in the magnetic field about a particular value $B$. $J(t|B)$ is the expected intensity at time $t$ given the magnetic field is $B$ and $\Delta B = 0.05 \ \mathrm{G}$ is the step size between simulated magnetic fields. Summing the expression over all magnetic field strengths in the study, we obtain the heuristic
\begin{equation}\label{eq:classical_Fisherinfo_broadband}
    h(t) = \sum_B \frac{1}{J(t|B)} \left ( \frac{J(t|B + \Delta B) - J(t|B)}{\Delta B} \right )^2,
\end{equation}
which describes how sensitive the signal is to small changes in the magnetic field on average given each magnetic field strength in the range is equally likely. Figure \ref{fig:CFI_broadbandaddresing} shows $h(t)$. 

Most of the variation of the signal indeed occurs between $10$~and~$11~\mu\mathrm{s}$. Since this is simply a heuristic and small deviations are still occurring outside of this window, we propose that the experiment collects light on either side of this signal for a total duration of $4~\mu\mathrm{s}$. Further refinement of the method that considers shorter total measurement time could increase the total sensitivity by a factor of around two if one only inspects the main $1~\mu\mathrm
s$ interval or by even more if one finds an optimal tradeoff between even shorter time intervals and information loss.

\subsection{Neural Network Model and Training Details}
\label{app:NN model}
The MLP model consisted of 4 linear layers with a hidden size of 1024, a leaky ReLU activation between each layer, and a sigmoid activation for the final output. Skip connections were added by summing the input and output of each linear layer.

The 1D convolutional networks were comprised of 1D ResNet-type blocks, where each block consisted of layers in the following order: 1D convolution, 1D batch normalization, activation, 1D convolution, 1D batch normalization, and a final activation. The kernel size of the convolutional layers was 3, and the stride was 2 for the first layer and 1 for the second. The input to the block is added to the output of the second batch normalization layer, and combined before the last activation. The output of the last ResNet block is flattened and passed through a linear layer which predicts a single output. We varied the depth of these 1D ResNets from 3 – 6, and in all cases the model failed to learn.

The LSTM (GRU) models consisted of a 3 layer bidirectional LSTM (GRU) with a hidden size of 128, followed by a leaky ReLU activation, and finally by a linear layer to predict the scalar output.

The NODE model uses a single layer to project the input time traces into a latent space that serves at the initial state of the NODE layer. The NODE uses an MLP to parameterize the vector field, and is propagated over an arbitrary timespan of 0 to 1. The final timestep of the NODE layer is then passed through another MLP decoder to obtain a single scalar value for the prediction of B.

All models were trained using the PyTorch learning rate scheduler ‘Reduce Learning Rate on Plateau’ (RLROP), with a patience of 50 epochs, a reduction factor of 0.5, and a minimum learning rate of $10^{-10}$. Notably, we observed that some models required high initial learning rates ($10^{-2}$) to achieve any learning, or else they would predict the mean of the labels. Tuning the patience parameter was crucial for achieving the lowest possible error. We made use of the Adam optimizer for all models and a weight decay of $10^{-5}$ for regularization (although we observed no overfitting).

We tuned hyperparameters for activation functions, hidden layer sizes, and number of layers for the MLPs used as the vector field and decoder, initial learning rate, and RLROP patience using the Weights $\And$ Biases package.


\end{document}